\documentclass[preprint,aps,draft]{revtex4}
\usepackage{amsmath}
\usepackage{amssymb}
\usepackage{graphicx}% Include figure files
\usepackage{dcolumn}% Align table columns on decimal point
\usepackage{bm}% bold math

\begin{document}

\title{Torsion fields generated by the quantum effects of macro-bodies}
% Force line breaks with \\
\author{Da-Ming Chen$^{1,2}$} \email{cdm@nao.cas.cn}
%\author{Konstantin Zioutas$^{2}$} \author{Steen H. Hansen$^1$}
%\author{Kristian Pedersen$^1$} \author{H\aa kon Dahle$^{1,3}$}
%\author{Anastasios Liolios$^4$}
 \affiliation{ $^1$National Astronomical Observatories, Chinese Academy of Sciences, Beijing 100101, China;\\
$^2$School of Astronomy and Space Science, University of Chinese Academy of Sciences, Beijing
100049, China}

\begin{abstract}
  We generalize Einstein's General Relativity (GR) by assuming that all matter (including macro-objects) has quantum effects. An appropriate theory to fulfill this task is Gauge Theory Gravity (GTG) developed by the Cambridge group. GTG is a ``spin-torsion" theory, according to which, gravitational effects are described by a pair of gauge fields defined over a flat Minkowski background spacetime. The matter content is completely described by the Dirac spinor field, and the quantum effects of matter are identified as the spin tensor derived from the spinor field. The existence of the spin of matter results in the torsion field defined over spacetime. Torsion field plays the role of Bohmian quantum potential which turns out to be a kind of repulsive force as opposed to the gravitational potential which is attractive. The equivalence principle remains and essential in this theory so that GR is relegated to a locally approximate theory wherein the quantum effects (torsion) are negligible. As a toy model, we assume that the macro matter content can be described by the covariant Dirac equation and  apply this theory to the simplest radially symmetric and static gravitational systems. Consequently, by virtue of the cosmological principle, we are led to a static universe model in which the Hubble redshifts arise from the torsion fields.
\end{abstract}


\maketitle
\section{Introduction}
In physics, when contradictions between the predictions of a fundamental theory and the credible experimental observations are unavoidable, one may  modify the theory or assume some new types of matter, and both approaches must be verified by experiments.  It is now well established that when Einstein's General Relativity (GR) is applied to galaxies (always in Newtonian form) we need some mysterious matter called dark matter, and when applied to cosmology we need even more mysterious element called dark energy in addition to the dark matter. While more and more experimental instruments are constructed to probe the direct or indirect evidences of dark matter and dark energy, it is equally important for physicists to investigate the theories of gravity alternative to GR. In such a theory, we hope that there is no need to assume the existence of dark matter or dark energy or both. For example, it was shown by Lovelock~\citep{lovelock1971einstein,Lovelock1972} that a polynomial form of the Lagrangian is possible leading to higher order curvature corrections to GR known as the Lovelock theory of gravity. Specifically, second order Lovelock or Einstein-Gauss-Bonnet gravity has been investigated recently~\citep{PhysRevLett.55.2656,2005GReGr..37.1869K,Oikonomou_2021,brassel2022charged}. Another important extension to GR was achieved by constructing the so-called nonlinear Lagrangians, which are some arbitrary but well-defined functions of the Ricci scalar~\citep{buchdahl1970non}, and now this theory has been developed into the well-known $f(R)$ gravity (e.g., \cite{goswami2014collapsing}). In this paper, however, we attempt to extend GR in a quite different way.  Note the fact that GR has been tested with high precision only for the gravitational systems with spatial size equal to or smaller than the solar system (e.g., compact objects) when gravity is relatively strong. Therefore, it is possible that the astronomical observations for larger gravitational systems (galaxies, cluster of galaxies until the whole universe) may suggest that GR should be extended or modified when the gravitational field is weak (except for the very early universe). We first assume that macro bodies have observable quantum effects on large scales and then attempt to extend GR to describe them (which are usually believed to be negligibly small in actual practice), and we require that GR is recovered on small scales.

Up to date, it is well known that quantum effects are believed to be the exclusive properties for microparticles which are the subjects of quantum mechanics and quantum field theory. When gravity is present, there are so-called quantum gravity and quantum field theory in curved spacetime, and all such theories are significant only under the extreme conditions like very strong gravity and very high energy, as in the cases when we study the phenomena near the horizons of black holes and in very early universe. Although all macro bodies come down to microparticles as their final compositions, physicists believe that the quantum effects of macro bodies are ``averaged out" and thus negligibly small compared with other kinematic quantities. Note, however, that the preceding conclusions are based on an implicit assumption that the quantum effects of macro bodies arise only from their micro constituents, while the macro bodies themselves remain classical. We thus ask a question: what if we assume that the quantum effects of macro bodies arise from their own quantum randomness which are independent of their micro constituents? In order to understand our assumption, we need to know what the quantum effects of microparticles may imply. The quantum effects of electrons, for example, can be attributed to their quantum randomness (or called intrinsic randomness), this distinguishes remarkably from classical randomness (or called apparent randomness) of which the  Brownian motions is a typical example. It is well known that the random motions of pollen grains in water result from the frequent collisions of the molecules of water according to Newtonian dynamics, while quantum random motion is the intrinsic property of a particle itself that cannot be attributed to any external reasons and thus Newtonian dynamics fails to describe.  Our assumption is that, like electrons,  all macro bodies (including galaxies if we treat them as ``particles") have the quantum random motions and thus should exhibit quantum effects.

An immediate doubt about our assumption is: why we have never observed such quantum effects of macro bodies? The answer is simple: if the macro bodies and an observer (also an ``macro body") are located near the same (but arbitrary) point of spacetime manifold, then they have approximately the same randomness (due to gravity) and quantum effects cannot be observed. The locally ``classical" behavior of macro bodies is a familiar scientific fact, however, this fact does not necessarily require the macro bodies to behave classically on arbitrary large scales. It is believed that gravity makes disorders order, and order is fundamental for space and time and spacetime, or as we can say, spacetime is nothing else but continuous order of events. Thus no gravity, no spacetime. In any local inertial frames, the gravity is still present but the net gravity force is canceled by inertial force, so order or flat spacetime remains.  In special relativity, rigid rods are employed to form coordinate lattices so that any event has a position at any given time. According to GR, the global structure of curved spacetime can still be described by arbitrary curved rigid rods (depending on the matter distributions) with arbitrary length. However, quantum mechanics tells us that microparticles can escape the order depicted by the spacetime and thus get out of control of gravity. Therefore, if we extend the quantum randomness owned by microparticles to macro bodies, gravity of one body (or system) cannot make remote bodies in order. So we should give up the averaged background curved spacetime determined by all matter in the universe. Instead, we assume a spacetime formed by sewing together infinitely many local pieces of spacetime depicted by GR. That is to say, if the observed macro bodies and an observer are separated by a finite spatial distance, the quantum effects appear, and the strength of the quantum effects increases with the increasing distances between them. So the rods are rigid only approximately on small scales (i.e., rigid locally), but are flexible on large scales.  Consequently, GR is correct only approximately in a sufficiently small local region around any point of the spacetime manifold, wherein quantum effects can be ignored. For large systems, quantum effects appear among different parts, and thus GR should be replaced by a new theory of gravity that can account for them. As a summary, we aim to extend GR to a theory of gravity based on the following hypotheses:

1. Equivalence principle:  frames of reference undergoing acceleration and frames of reference in gravitational fields are equivalent.

2. The principle of general covariance: any physical predictions are independent of the frames of references employed, i.e, any physical theory can be written in a covariant form.

3. All matter has quantum random motion irrespective of its mass.

Clearly, compared with GR, the only added hypothesis is the quantum effects of macro matter, since the quantum effects of micromatter have been verified and well known. At the present stage, we emphasize that our extra hypothesis is compatible with Einstein's equivalence principle in the sense that, at any local point, a freefall  macro body has no quantum effects at its own rest inertial frame, so that its trajectory is well-defined locally.

We cannot go further until we have an optimal mathematical language to describe our assumptions in a natural way. We find that geometric algebra (GA) is just what we seek. GA is rediscovered and greatly promoted by David Hestenes \citep{1966STA+Hestenes,1986Hestenes+Sobczyk,2002.book.new.foundations.Hestenes}, and among other things, it provides a unified mathematical language for relativity (both special and general theories) and quantum mechanics~\citep{2003AmJPh..71..104H,2003AmJPh..71..691H,doran2003geometric}. We summarize the main results of spacetime algebra and Dirac theory for free particles in Appendix~\ref{sec:STA and Dirac}. Fortunately, we have a theory of gravity based on GA as we need at hand. More than twenty years ago, a new gauge theory of gravity constructed on flat spacetime alternative to GR was developed by Cambridge  group (Lasenby, Doran, and Gull) called gauge theory gravity (GTG)~\citep{1998RSPTA.356..487L}. When applying GTG for our purpose, we still use the Dirac spinor to describe the macro matter content. The only modification to GTG is that, when gravity is absent, we require that the gauge invariant equations for Dirac spinor are reduced to a conservation law for ``classical " (i.e., no spin) pressure free ideal fluid, rather than to the usual Dirac equation in flat spacetime (which is required for electrons). Unfortunately, up to date, we do not know the spinor field equations that macro bodies may satisfy. So as a first try, we still employ the Dirac equation for free particles as a toy model for our macro bodies. We outline the main idea and the prime processes followed by GTG for the matter content that can be described by Dirac equation in Appendix~\ref{sec:GTG}.

In what follows, we first apply GTG to cosmology and the brief conclusions and discussions are given at the end of the paper. We employ natural units ($G=\hslash=c=1$) throughout except stated otherwise.

\section{Cosmology: static universe model and Hubble redshift}
The simplest application of our theory that can be tested with credible observational data is the study of cosmology. The standard $\Lambda$CDM cosmology model is based on GR, in which the universe is expanding (accelerating at the present stage), the quantum effects are important only in very early universe~\citep{1997GReGr..29.1527C}. However, according to the assumption in this paper, macro bodies also have quantum effects, we expect that the quantum pressure (torsion field, or spin field) will balance the gravity, so that the whole universe is in a state of equilibrium all the time. So we should try a static universe model, in which $\bar{h}$ field is independent of time, and the average matter density $\rho_m=Nm\rho$ is a constant parameter, where $N$ is the number of macro bodies of mass $m$ for a finite system and is suppressed in our subsequent calculations. Thus the Dirac spinor $\psi(r)$ is a function of $r$ only, although the probability density $\rho=\psi\tilde{\psi}$ is also a constant due to the constant value of $\rho_m$. As for the magnitude of the spin density $S$, we assume that it is proportional to $\rho_m$, and since it has dimension of $\hbar$, we simply take the proportionality coefficient to be unity to match the covariant Dirac equation and the matter stress-energy tensor (see below).

We start by defining a set of spherical coordinates. From the position vector of the flat spacetime
\begin{equation}\label{eq:position vector}
x=t\gamma_0+r\sin\theta(\cos\phi\gamma_1+\sin\phi\gamma_2)+r\cos\theta\gamma_3
\end{equation}
we obtain
\begin{equation}\label{eq:spherical basis}
\begin{split}
  e_t=\partial_t x=& \gamma_0, \\
  e_r=\partial_r x=&\sin\theta(\cos\phi\gamma_1+\sin\phi\gamma_2)+\cos\theta\gamma_3,  \\
  e_\theta=\partial_\theta x=&r\cos\theta(\cos\phi\gamma_1+\sin\phi\gamma_2)-r\sin\theta\gamma_3, \\
  e_\phi=\partial_\phi x=&r\sin\theta(-\sin\phi\gamma_1+\cos\phi\gamma_2).
\end{split}
\end{equation}
From these vectors, the corresponding reciprocal basis {$e^t, e^r, e^{\theta}, e^{\phi}$} can be easily obtained. Since $e_\theta$ and $e_\phi$ are not unit, we define
\begin{equation}\label{eq:unit basis}
\hat{\theta}\equiv e_\theta/r, \; \; \; \hat{\phi}\equiv e_\phi/(r\sin\theta).
\end{equation}
With these unit vectors, we further define the unit bivectors (relative basis vectors for $e_t=\gamma_0$)
\begin{equation}\label{eq:pauli basis}
\begin{split}
   \sigma_r\equiv &e_re_t, \\
     \sigma_\theta\equiv &\hat{\theta}e_t, \\
     \sigma_\phi\equiv &\hat{\phi}e_t.
\end{split}
\end{equation}
These bivectors satisfy
\begin{equation}\label{eq:pseudo-scalar}
\sigma_r\sigma_\theta\sigma_\phi=e_te_r\hat{\theta}\hat{\phi}=i.
\end{equation}

For our purpose, we next try a form of $\bar{h}$ field that satisfies a static, radially symmetric matter distribution. We assume that it takes the form~\citep{1998RSPTA.356..487L,doran2003geometric}
\begin{equation}\label{eq:h function}
\begin{aligned}
   &\bar{h}(e^t)=f_1e^t+f_2e^r,   &&\bar{h}(e^r)=g_1e^r+g_2e^t,  \\
   &\bar{h}(e^{\theta})=\alpha e^{\theta},  &&\bar{h}(e^{\phi})=\alpha e^{\phi},
\end{aligned}
\end{equation}
where $f_1$, $f_2$, $g_1$, $g_2$ and $\alpha$ are all functions of $r$ only. From (\ref{eq:omega}) and (\ref{eq:H function}), we write $\omega'(a)$ as~\citep{1998RSPTA.356..487L}
\begin{equation}\label{eq:omega prime as a function of r}
\begin{aligned}
\omega'(a)=&(a\cdot e_t G-a\cdot e_r F)e_re_t-a\cdot\hat{\theta}X\hat{\theta}e_t \\
 & -a\cdot\hat{\theta}(Y-\frac{\alpha}{r})e_r\hat{\theta}-a\cdot\hat{\phi}X\hat{\phi}e_t \\
 & -a\cdot\hat{\phi}(Y-\frac{\alpha}{r})e_r\hat{\phi},
\end{aligned}
\end{equation}
where new functions $G$, $F$, $X$ and $Y$ are also all functions of $r$ only, their introduction circumvents the unnecessary complexity when derived directly in terms of the functions $f_1$, $f_2$ etc. With the $\omega$ function at hand, the curvature tensor can be calculated directly from (\ref{eq:curvature}), the torsion free part is ~\citep{1998RSPTA.356..487L}
\begin{equation}\label{eq:R'B}
\begin{split}
   R'(B)= & \alpha_1\sigma_r B\cdot\sigma_r+(\alpha_2\sigma_\theta+\alpha_3i\sigma_\phi)B\cdot\sigma_\theta \\
     & +(\alpha_2\sigma_\phi-\alpha_3i\sigma_\theta)B\cdot\sigma_\phi+\alpha_6\sigma_r(B\wedge\sigma_r) \\
     & +(\alpha_4\sigma_\theta-\alpha_5i\sigma_\phi)(B\wedge\sigma_\theta) \\
     & +(\alpha_4\sigma_\phi+\alpha_5i\sigma_\theta)(B\wedge\sigma_\phi),
\end{split}
\end{equation}
where $B$ is a bivector, and $\alpha_1, \ldots, \alpha_6$ are given by
\begin{equation}\label{eq:alphas}
\begin{array}{lll}
  \alpha_1=L_rG-L_tF+G^2-F^2, & \alpha_2=-L_tX+GY-X^2,  \\
  \alpha_3=L_tY+XY-XG, & \alpha_4=L_rY+Y^2-FX, \\
  \alpha_5=L_rX+XY-FY, & \alpha_6=-X^2+Y^2-(\frac{\alpha}{r})^2.
\end{array}
\end{equation}
From (\ref{eq:R'B}), the torsion free part of Ricci tensor and Ricci scalar are given by
\begin{equation}\label{eq:R'a}
\begin{split}
   R'(a)= & [(\alpha_1+2\alpha_2)a\cdot e_t+2\alpha_5a\cdot e_r]e_t \\
     & +[2\alpha_3a\cdot e_t-(\alpha_1+2\alpha_4)a\cdot e_r]e_r \\
     & -(\alpha_2+\alpha_4+\alpha_6)a\cdot\hat{\theta}\hat{\theta} \\
     & -(\alpha_2+\alpha_4+\alpha_6)a\cdot\hat{\phi}\hat{\phi},
\end{split}
\end{equation}
\begin{equation}\label{eq:R'}
R'=2\alpha_1+4\alpha_2+4\alpha_4+2\alpha_6.
\end{equation}

In order to solve Einstein equation (\ref{eq:Einstein equation}), we need write out $T(a)$ given by (\ref{eq:energy-momentum tensor}) explicitly, which is
\begin{equation}\label{eq:energy-momentum tensor 2}
\begin{split}
   T(a)&=\langle a\cdot D\psi i\gamma_3\tilde{\psi}\rangle_1 \\
     &=\langle a\cdot\bar{h}(e^{\mu})\partial_{\mu}\psi i\gamma_3\tilde{\psi}+\frac{1}{2}\omega(a)\psi i\gamma_3\tilde{\psi}\rangle_1 \\
     &=a\cdot(g_1e^r+g_2e^t)\langle\partial_r\psi i\gamma_3\tilde{\psi}\rangle_1+\omega(a)\cdot S.
\end{split}
\end{equation}
The term $\langle\partial_r\psi i\gamma_3\tilde{\psi}\rangle_1$ in the last equation can be calculated straightforward from Dirac equation (\ref{eq:Dirac equation}). We obtain
\begin{equation}\label{eq:energy-momentum tensor 3}
\begin{split}
   T(a)= & \frac{g_2a\cdot e_t-g_1a\cdot e_r}{g_1^2-g_2^2}\{(\rho_m-\frac{3}{2}\kappa S^2)(g_1e_r-g_2e_t) \\
     & +[g_2(G+2(Y-\frac{\alpha}{r}))-g_1(F+2X)](e_re_t)\cdot S\}.
\end{split}
\end{equation}
Now set $a=e_t$ and $a=e_r$ in (\ref{eq:Einstein equation}) respectively, we obtain the following set of equations:
\begin{equation}\label{eq:equation for alphas}
\begin{split}
   2\alpha_4+\alpha_6+\frac{3}{4}\kappa^2S^2= & \frac{g_2^2}{g_1^2-g_2^2}\kappa(\rho_m-\frac{3}{2}\kappa S^2) \\
   \alpha_3=\alpha_5= & \frac{g_1g_2}{2(g_1^2-g_2^2)}\kappa(\rho_m-\frac{3}{2}\kappa S^2) \\
   2\alpha_2+\alpha_6+\frac{3}{4}\kappa^2S^2= & -\frac{g_1^2}{g_1^2-g_2^2}\kappa(\rho_m-\frac{3}{2}\kappa S^2).
\end{split}
\end{equation}

Recall that the functions $G(r)$, $F(r)$, $X(r)$ and $Y(r)$ are introduced only for the torsion-free parts $\omega'$ of $\omega$ function, they thus should be related to the functions $f_1$ etc in $\bar{h}$ field under the same conditions (i.e., torsion free). Note also that $L_t=e_t\cdot\hat{h}(e^{\mu})\partial_{\mu}=g_2\partial_r$ and $L_r=e_r\cdot\hat{h}(e^{\mu})\partial_{\mu}=g_1\partial_r$, we further adopt the suggestions in~\cite{1998RSPTA.356..487L} that $f_2=0$, $\alpha=1$, $g_1=L_r r=Yr$ and $g_2=L_t r=Xr$. Applying all these relations to the equations in (\ref{eq:equation for alphas}), and defining $M_Q=-\frac{1}{2}r^3\alpha_6=\frac{r}{2}(g_2^2-g_1^2+1)$, we find
\begin{equation}\label{eq:LrM}
L_r M_Q=\frac{3}{8}g_1\kappa^2S^2r^2, \quad \text{or} \quad  \partial_rM_Q=\frac{3}{8}\kappa^2S^2r^2.
\end{equation}
Since $S=\frac{1}{2}\psi i\gamma_3\tilde{\psi}$ and due to our special choice of $\psi=\rho^{1/2}R$, we have $S^2=-\frac{1}{4}\rho^2<0$. For a static homogeneous universe with constant matter density $\rho_m =\rho m$, $S^2$ is also a minus constant. By integrating (\ref{eq:LrM}) we get
\begin{equation}\label{eq:MQ}
M_Q=\frac{1}{8}\kappa^2S^2r^3=\frac{r}{2}(g_2^2-g_1^2+1),
\end{equation}
where we have set the integration constant to zero since we require $M_Q=0$ when $r=0$. Remarkably, if we interpret $M_Q$ as some kind of ``mass" within $r$, then it provides a repulsive force! In our gauge choice, we can define a Bohmian quantum potential $\Phi_Q(r)=\frac{M_Q(r)}{r}=\frac{1}{8}\kappa^2S^2r^2$. To keep the whole universe static, this potential should balance the usual matter gravitational potential $\Phi_m(r)=\frac{4\pi\rho_m}{3}r^2$. We thus have
\begin{equation}\label{eq:rhom2rhoQ}
\rho_m=-\frac{3}{4}\kappa S^2.
\end{equation}

As mentioned, we should relate $G$ to the functions $f_1$, $g_1$ in $\bar{h}$ field under torsion-free condition. In this case, the torsion equation (\ref{eq:torsion}) is reduced to the ``Wedge equation" $D'\wedge\bar{h}(a)=0$. Let $a=e_t$ in this equation we obtain (the same result is obtained by~\cite{1998RSPTA.356..487L} in a different way)
\begin{equation}\label{eq:f1 to G}
f_1=e^{-\int\frac{G}{g_1}dr}.
\end{equation}
From (\ref{eq:equation for alphas}), we get (assuming $g_2=0$)
\begin{equation}\label{eq:G/g1}
-\frac{G}{g_1}=\frac{\kappa(\rho_m-\kappa S^2)r}{2(1-\kappa^2S^2r^2/4)}.
\end{equation}
Substituting this expression into (\ref{eq:f1 to G}) and applying (\ref{eq:rhom2rhoQ}) we get
\begin{equation}\label{eq:f1}
\begin{split}
f_1(r)&=\left(1-\frac{1}{4}\kappa^2S^2r^2\right)^{\frac{\rho_m-\kappa S^2}{-\kappa S^2}} \\
      &=\left(1+H_0^2r^2\right)^{\frac{7}{4}},
\end{split}
\end{equation}
where
\begin{equation}\label{eq:Hubble parameter}
H_0^2=\frac{8\pi\rho_m}{3}
\end{equation}
is the Hubble constant. We are now ready to study the redshifts of light signals emitted from some source at distance $r$. According to the equivalence principle, the line element of events given by (\ref{eq:line element}) is still valid in our theory. From (\ref{eq:square root of metric}), $g^t=\bar{h}(e^t)=f_1e^t$, we thus get $g_t=\underline{h}^{-1}(e_t)=f_1^{-1}e_t$. Then from (\ref{eq:line element}), the period of the light signal at the source is
\begin{equation}\label{eq:period}
\Delta t(r)=f_1\Delta\tau=\left(1+H_0^2r^2\right)^{\frac{7}{4}}\Delta\tau,
\end{equation}
where $\Delta\tau$ is the invariant proper value of the period. So the redshift we observed is
\begin{equation}\label{eq:redshift}
1+z(r)=\frac{\Delta t(r)}{\Delta t(0)}=\left(1+H_0^2r^2\right)^{\frac{7}{4}}.
\end{equation}

Clearly, the redshifts derived in this manner arise purely from the time dilations induced by the ``quantum potential" $\Phi_Q(r)=\frac{1}{8}\kappa^2S^2r^2$ rather than the gravitational potential or expanding of the universe, since the matter distribution is homogeneous and the whole universe is static. According to quantum mechanics, however, the result we obtained here is not a surprising. As indicated by David Hestenes~\citep{hestenes1997real}, when negative muons are captured in atomic $s$-states their lifetimes are increased by a time dilation factor corresponding to the Bohr velocity. Clearly, such a time dilation results from quantum potential, or quantum pressure or zero-point energy so to speak.  We emphasize that all the quantum effects of matter are the same irrespective of the mass of the matter (i.e., no matter the matter is macro or micro). So an observer located at any place in the universe would observe the redshifted light signals emitted from the sources located far away from the observer in arbitrary directions. This explains the observations of Hubble redshifts. It should be also pointed out that the concepts of quantum potential or quantum force originally proposed by David Bohm correspond to the ``quantum pressure" or ``zero-point energy" usually referred to by the major physicists according to their ``standard" viewpoint of quantum mechanics. It is not helpful for us to indulge in the controversies between the two different interpretations of quantum mechanics, it suffices to know the fact that the causal or particle (trajectories or histories) interpretations of quantum effects of matter are the very natural results if one employs GA as the mathematical language: the spinor $\psi=\rho^{\frac{1}{2}}R$ determines a unique family of matter trajectories and all other conclusions follow.

\section{Summary and discussion}
We attempt to generalize Einstein's GR by adding the third hypothesis concerning the quantum effects of macro bodies to form a new theory of gravity. This must be a ``spin-torsion" theory of gravity, except that we require the torsion field (or the spin of source matter) to vanish locally according to the equivalence principle when only macro matter is involved. GTG is such a theory in the sense that it can define the gravitational strength corresponding to $\omega$ field self-consistently both from a Dirac spinor field $\psi(x)$ or a multivector field $M(x)$. Consequently, the obtained curvature tensor $R(a\wedge b)$ may or may not contain the torsion depending on whether or not the source matter having spins. Employing spinor field in the form $\psi(x)=\rho^{\frac{1}{2}}(x)R(x)$ to represent the macro matter distributions is essential to our theory. The reason is that this form of spinor, obtained by setting $\beta=0$ in the more general canonical form $\psi=(\rho e^{i\beta})^{1/2}R$, naturally ensures the neutral matter current in the free falling frames. Although GTG incorporates quantum effects in gravity only for microparticles, and the minimal coupling procedure (gauge principle, which results in or replaces Einstein's equivalence principle) ensures that the minimally coupled Dirac action yield the minimally coupled Dirac equation, one can manage to derive some spinor equations such that torsion vanishes naturally in an approximate way in any sufficiently small local regions. This is equivalent to say that the quantum spin of macro matter cannot be observed locally. Therefore, in our new theory of gravity, GR is valid only approximately in local regions or on small scales.

Before finding out a correct spinor equation for macro matter , we are eager to know the possible results for practical applications. Dirac equation has been well-studied when gravity is present, although we know that it does not meet our demand.  As a toy model, we have employed the covariant Dirac equation for free particles to describe the spherically symmetric and static gravitational systems. Remarkably, we find that such a system has a negative mass determined by the spin density. We then applied the results to cosmology and find that the repulsive force provided by the negative mass can balance the gravity and we achieve a static universe model. We have arrived at an expression for the Hubble redshift as a function of the distance to the light source. Interestingly, from this expression we can take the constant average density of the universe $\rho_m$ as a fundamental physical constant (like gravitational constant $G$, the speed of light $c$, etc). The disadvantages of the toy model are obvious. The procedure we followed to define the mass of the system is similar to the torsion-free ideal fluid model, however, the obtained mass is negative, which is nothing else but the spin of the system. Surprisingly, the real (positive) mass is ``lost" in our calculations, so the balance condition (\ref{eq:rhom2rhoQ}) has to be put by hand! Further more, for $z\ll 1$, our formula gives $z\sim r^2$, which is not linear of distance $r$ as usually declared by astronomers. These disadvantages are the evidence that Dirac equation is not the correct one we are seeking.

From a spinor field we can always form vector currents $\rho e_{\mu}=\psi\gamma_{\mu}\tilde{\psi}$. In any local free-falling frames, we require these currents to be constant, so that if we identify $e_3=R\gamma_3\tilde{R}$ as the spin axis vector $s/|s|$ and $e_1\wedge e_2$ as the spin plane, then the angular velocity in the spin plane is zero, and thus no spin can be observed, which is just what we need for macro matter. This may suggest a equation $\nabla\psi=0$, but its solutions do not satisfy the Dirac equation unless $m=0$. This can partly explain why we cannot employ the Dirac equation directly in our new theory. On the other hand, when gravity is present, covariant derivative of spinor fields would produce non-zero spin or torsion fields. Like the torsion-free Riemann curvature fields, torsion fields constructed in this way would ensure that the redshift or relative time dilation originated from macro quantum effects increases with increasing distances between observers and light sources.

%\normalem
\begin{acknowledgements}
The author is very grateful to the anonymous referee for the good evaluation of the paper and the helpful suggestions for improvements.  This work is supported by the NSFC grant (No. 11988101) and the K.C.Wong Education Foundation. The author would like to thank Liang Gao in NAOC for his strong financial support for many years.
\end{acknowledgements}

\bibliographystyle{apsrev4-1}
\bibliography{dmchenrf}

\appendix

\section{Spacetime algebra and Dirac theory}\label{sec:STA and Dirac}
The geometric algebra (GA) that is generated by a 4-dimensional Minkowski vector space is called spacetime algebra (STA). The inner and outer products of the four orthonormal basis vectors in Minkowski vector space \{$\gamma_{\mu}, \mu=0 \ldots 3$\} are defined to be
\begin{eqnarray}\label{eq:dot and wedge products for basis}
  \gamma_{\mu}\cdot\gamma_{\nu}\equiv\frac{1}{2}(\gamma_{\mu}\gamma_{\nu}+\gamma_{\nu}\gamma_{\mu}) &\equiv & \eta_{\mu\nu}=\text{diag}(+ - - -) \nonumber \\
  \gamma_{\mu}\wedge\gamma_{\nu} &\equiv &\frac{1}{2}(\gamma_{\mu}\gamma_{\nu}-\gamma_{\nu}\gamma_{\mu}).
\end{eqnarray}
A full basis for the STA is
\begin{equation}\label{eq:full set of STA}
1, \; \{\gamma_{\mu}\}, \; \{\sigma_{k}, i\sigma_{k}\}, \; \{i\gamma_{\mu}\}, \; i
\end{equation}
where $\sigma_k\equiv\gamma_k\gamma_0, k=1\ldots 3$, and $i=\gamma_0\gamma_1\gamma_2\gamma_3=\sigma_1\sigma_2\sigma_3$. The STA is a linear space of dimension $1+4+6+4+1=2^4=16$. We call the general elements of STA multivectors, and each multivector decomposes into a sum of elements of different grades. The grade-$r$ multivectors $A$ are denoted by $A_r$. We call grade-$0$ multivectors scalars, grade-$1$ vectors, grade-$2$ bivectors and grade-$3$ trivectors. The geometric product of a grade-$r$ multivector $A_r$ with a grade-$s$ multivector $B_s$ is defined simply by $A_rB_s$, which decomposes into
\begin{equation}\label{eq:GA product}
A_rB_s=\langle A_rB_s\rangle_{|r-s|}+\langle A_rB_s\rangle_{|r-s|+2}+...+\langle A_rB_s\rangle_{r+s},
\end{equation}
where $\langle X\rangle_r$ denotes the projection onto the grade-$r$ part of $X$. The grade-$0$ (scalar) part of $X$ is written $\langle X\rangle$.
We employ ``$\cdot$" and ``$\wedge$" symbols to denote the lowest-grade and highest-grade terms in (\ref{eq:GA product}), so that
\begin{eqnarray}\label{eq:dot and wedge product}
  A_r\cdot B_s &=& \langle A_rB_s\rangle_{|r-s|},\label{eq:dot product Ar Bs} \\
  A_r\wedge B_s &=& \langle A_rB_s\rangle_{r+s},
\end{eqnarray}
which are called inner and outer products respectively. The simple example is the geometric product of two vectors $a$ and $b$
\begin{equation}\label{eq:GA product ab}
ab=a\cdot b+a\wedge b.
\end{equation}

We define the reverse of a geometric product $AB$ by $(AB)^{\sim}=\tilde{B}\tilde{A}$, so that for vectors $a_1, a_2, ..., a_r$, we have $\tilde{a}_1=a_1$ and
\begin{equation}\label{eq:reverse}
(a_1a_2...a_r)^{\sim}=a_ra_{r-1}...a_1.
\end{equation}
It is easy to show that
\begin{equation}\label{eq:reverse Ar}
 \tilde{A}_r=(-1)^{r(r-1)/2}A_r.
\end{equation}
Thus suppose $r\leq s$, the inner and outer products satisfy the symmetry properties
\begin{eqnarray}
  A_r\cdot B_s &=& (-1)^{r(s-1)}B_s\cdot A_r, \\
 A_r\wedge B_s &=& (-1)^{rs}B_s\wedge A_r.
\end{eqnarray}
The scalar product is defined by
\begin{equation}\label{eq:scalar product}
A\ast B=\langle AB\rangle.
\end{equation}
From (\ref{eq:dot product Ar Bs}), $A_r\ast B_s$ is nonzero only if $r=s$, thus the scalar product (\ref{eq:scalar product}) is commutative
\begin{equation}\label{eq:commutative scalar product}
\langle AB\rangle=\langle BA\rangle.
\end{equation}
This commutative property of the scalar product is very useful in later calculations. We further define the commutator product
\begin{equation}\label{eq:commutator product}
A\times B=\frac{1}{2}(AB-BA),
\end{equation}
which satisfies the Jacobi identity
\begin{equation}\label{eq:Jocobi}
A\times(B\times C)+B\times(C\times A)+C\times(A\times B)=0.
\end{equation}

The advantages of STA are that it enables coordinate-free representation and computation of physical systems and processes, and it incorporates the spinors of quantum mechanics along with the tensors of classical field theory.

Geometric calculus (GC) is the extension of a geometric algebra (like STA) to include differentiation and integration. Let multivector $F$ be an arbitrary function of some multivector argument $X$, then the derivative of $F(X)$ with respective to $X$ in the $A$ direction is defined by
\begin{equation}\label{eq:multivector directional derivative}
A\ast\partial_{X}F(X)\equiv\lim_{\tau\rightarrow 0}\frac{F(X+\tau A)-F(X)}{\tau},
\end{equation}
where the multivector partial derivative $\partial_{X}$ inherits the multivector properties of its argument $X$. We have
\begin{equation}\label{eq:partial derivative for scalar functions}
\partial_{X}\langle XA\rangle=P_{X}(A),
\end{equation}
where $P_{X}(A)$ is the projection of $A$ onto the grades contained in $X$. For vector argument $x$ and constant vector $a$, (\ref{eq:multivector directional derivative}) and (\ref{eq:partial derivative for scalar functions}) give
\begin{equation}\label{eq:vector derivative for x}
a\cdot\partial_x x=a=\partial_x(x\cdot a).
\end{equation}
For a vector variable $a=a^{\mu}\gamma_{\mu}=a\cdot\gamma_{\mu}\gamma^{\mu}=a\cdot\gamma^{\mu}\gamma_{\mu}$, where $\gamma^{\mu}$ constitute the reciprocal basis and satisfy $\gamma_{\mu}\cdot\gamma^{\nu}=\delta_{\mu}^{\nu}$, the vector derivative can be defined as
\begin{equation}\label{eq:vector derivative a}
\partial_a\equiv\gamma^{\mu}\frac{\partial}{\partial a^{\mu}},
\end{equation}
For the derivative with respect to a spacetime position vector $x$ we use the symbol $\nabla\equiv\partial_x=\gamma^{\mu}\frac{\partial}{\partial x^{\mu}}$, if $x=x^{\mu}\gamma_{\mu}$. From (\ref{eq:vector derivative for x}) and (\ref{eq:vector derivative a}) we can obtain useful results
\begin{equation}\label{eq:vector partial derivative and basis}
\partial_a=\partial_{b}b\cdot\partial_a=\gamma^{\mu}\gamma_{\mu}\cdot\partial_a.
\end{equation}
Once again, one great advantage of GC is that it eliminates unnecessary conceptual barriers between classical, quantum and relativistic physics.

We now discuss the Dirac theory in terms of STA. Recall the symbols $\gamma's$ and $\sigma's$ in (\ref{eq:dot and wedge products for basis}) and (\ref{eq:full set of STA}), which stand for the basis vectors of 4-dimensional Minkowski space and the relative three-dimensional space, respectively. The same symbols are used in Dirac theory of relativistic quantum mechanics, but therein they are matrices ($4\times 4$ for Dirac $\gamma$'s and $2\times 2$ for Pauli $\sigma$'s). This is not a coincidence, since they satisfy exactly the same algebraic equations. In view of STA, this correspondence reveals the geometric properties of Dirac spinor, and leads to the causal interpretation of quantum mechanics (i.e., microparticles have the well-defined trajectories in spacetime), which are of prime importance for the present work. We call the elements of the even subalgebra of STA defined in spacetime as Dirac spinor fields, denoted by $\psi(x)$, and can be written in the canonical form
\begin{equation}\label{eq:spinor}
\psi(x)=(\rho e^{i\beta})^{1/2}R(x),
\end{equation}
where $\rho(x)$ is the proper probability density, $i=\gamma_0\gamma_1\gamma_2\gamma_3$, $\beta(x)$ is a scalar field and $R(x)$ (called a rotor) satisfies the normalization condition $R\tilde{R}=1$. The rotor $R(x)$ determines a Lorentz rotation of a given fixed frame vectors $\gamma_{\mu}$ into a frame $e_{\mu}$ given by
\begin{equation}\label{eq:Lorentz rotation}
e_{\mu}=R\gamma_{\mu}\tilde{R},
\end{equation}
and according to this we write
\begin{equation}\label{eq:curent}
\rho e_{\mu}=\psi\gamma_{\mu}\tilde{\psi}.
\end{equation}
We identify $e_0=v$ as the proper velocity of a particle of which the spinor (wave function) is $\psi$, so that $v^2=1$ and
\begin{equation}\label{eq:Dirac current}
\rho v=\rho e_0=\psi\gamma_0\tilde{\psi}
\end{equation}
is the Dirac current. We interpret another vector field as $e_3=R\gamma_3\tilde{R}=s/|s|$, where $|s|=\frac{\hslash}{2}$ is the magnitude of the spin vector $s$ (we recover the Planck constant $\hslash$ from unity in this Appendix). The spin angular momentum $S(x)$ is a bivector field related to $s(x)$ by
\begin{equation}\label{eq:spin angular momentum}
S=isv=\frac{\hslash}{2}ie_3e_0=\frac{\hslash}{2}Ri\sigma_3\tilde{R}=\frac{\hslash}{2}R\gamma_2\gamma_1\tilde{R}.
\end{equation}

In STA, the Dirac equation for a free-particle of mass $m$ is~\citep{1967JMP.....8..798H,1973JMP....14..893H}
\begin{equation}\label{eq:Dirac equation in flat spacetime}
\nabla\psi i\sigma_3=m\psi\gamma_0,
\end{equation}
which admits plane wave solutions of the form \citep{hestenes1997real}
\begin{equation}\label{eq:plane wave}
\psi(x)=(\rho e^{i\beta})^{\frac{1}{2}}R=(\rho e^{i\beta})^{\frac{1}{2}}R_0 e^{-i\sigma_3p\cdot x/\hslash},
\end{equation}
where $R_0$ is independent of position vector $x$ and $p$ is the momentum vector of the free particle, so the rotor $R$ has been decomposed to explicitly exhibit its spacetime dependent in a phase factor. Inserting this into (\ref{eq:Dirac equation in flat spacetime}) and using (\ref{eq:vector derivative for x}), we obtain
\begin{equation}\label{eq:p times psi}
p\psi=m\psi\gamma_0.
\end{equation}
Right multiplying by $\tilde{\psi}$ we get
\begin{equation}\label{eq: p from free Dirac}
p=me^{i\beta}R\gamma_0\tilde{R}=mve^{-i\beta},
\end{equation}
where we have used $i\gamma_0=-\gamma_0 i$ and $iR=Ri, i\tilde{R}=\tilde{R}i$. Since $e^{i\beta}=\cos\beta+i\sin\beta$, to ensure the momentum $p$ to be a vector, we must have $\beta=0$ or $\pi$. We can identify these as corresponding to the electron or positron wave functions (i.e., spinors) respectively. Since we want to use spinor field to describe neutral macro body, this result suggests that we should set $\beta=0$ in this paper. Thus for a free electron, $p=mv$, i.e., the momentum is collinear with the proper velocity, which in general is not true.

We now consider the expected trajectories (streamlines) of a free electron. The position vector is $x(\tau)$, where $\tau$ is the proper time, so we have
\begin{equation}\label{eq:proper velocity}
v=\frac{dx}{d\tau}=\dot{x}=e_0=R\gamma_0\tilde{R}.
\end{equation}
Then the spinor associated to the trajectory is
\begin{eqnarray}\label{eq:spinor for the trajectory}
\psi(\tau)&=&\rho^{\frac{1}{2}}R_0e^{-\gamma_2\gamma_1(mv\cdot x)/\hslash} \nonumber \\
&=&\rho^{\frac{1}{2}}R_0e^{-\gamma_2\gamma_1m\tau/\hslash} \nonumber \\
&=&\rho^{\frac{1}{2}}R_0e^{-\gamma_2\gamma_1\omega\tau/2},
\end{eqnarray}
where $\tau=v\cdot x$ and $\omega=2m/\hslash$ is the angular velocity in the spin plane $e_2e_1$. Since $\psi(\tau)=\rho^{1/2}R(\tau)$, (\ref{eq:spinor for the trajectory}) implies
\begin{equation}\label{eq:rotor for free trajectory}
R(\tau)=R_0e^{-\frac{1}{2}\gamma_2\gamma_1\omega\tau}.
\end{equation}
Clearly, the proper velocity of the free particle $v=R_0\gamma_0\tilde{R}_0$ and the spin vector $s=\frac{\hslash}{2}R_0\gamma_3\tilde{R}_0$ are constant, however, for $k=1,2$,
\begin{eqnarray}\label{eq:e1 and e2}
e_k(\tau)&=&R(\tau)\gamma_k\tilde{R}(\tau) \nonumber \\
&=&R_0e^{-\frac{1}{2}\gamma_2\gamma_1\omega\tau}\gamma_ke^{\frac{1}{2}\gamma_2\gamma_1\omega\tau}\tilde{R}_0 \nonumber \\
&=&R_0\gamma_ke^{\gamma_2\gamma_1\omega\tau}\tilde{R}_0 \nonumber \\
&=&R_0\gamma_k\tilde{R}_0R_0 e^{\gamma_2\gamma_1\omega\tau}\tilde{R}_0 \nonumber \\
&=&e_k(0)e^{e_2e_1\omega\tau},
\end{eqnarray}
where $e_k(0)=R_0\gamma_k\tilde{R}_0$. Thus, the expected trajectories are straight lines ($v$ is constant), and as the electron moves along the straight trajectory, the spin is constant and $e_1$ and $e_2$ rotate about $e_3$ (spin axis) with angular velocity $\omega$.

We conclude that, for free electrons, the probable trajectories $x(\tau)$ are well-defined, and along them, the proper velocity $v$ is constant and satisfies $v^2=1$ and the spin vector $s$ is also constant with magnitude $\frac{\hslash}{2}$. Further studies of this particle-interpretation of Dirac theory show us that, quantum effects can be determined by the spins of electrons, so that no spins, no quantum effects~\citep{2003AmJPh..71..691H}. Surprisingly, when gravity is present, Hestenes has applied the spinor approach to compute the gravitational precession of a gyroscope and found that gravitational effects on electron motion are exactly the same on the classical rigid body motion~\citep{1986IJTP...25..589H}.  Encouragingly, these studies are very close to our aim since we want to extend GR to describe quantum effects of macro bodies. So it is natural for us to choose Dirac spinor field to represent matter field
\begin{equation}\label{eq:matter field psi}
\psi(x)=\rho^{\frac{1}{2}}(x)R(x).
\end{equation}
According to our hypotheses, however, in any local inertial frames, quantum effects of macro bodies should vanish, so we cannot directly employ Dirac equation for our spinor fields. The classical spins such as that for gyroscopes will not be considered since they have nothing to do with quantum effects. So what we seek is a differential equation for the spinor of the form in (\ref{eq:matter field psi}), such that, in local inertial frames it describes a pressure-free (classical) ideal fluid, and when gravity is present quantum effects appear naturally. However, before finding out the correct equations, it would be interesting to apply the Dirac equation to a system of $N$ identical macro particles of mass $m$, so that matter density is $\rho_m=Nm\psi\tilde{\psi}$. We know that Dirac equation does not meet our demand, but as a toy model, we apply it to cosmology and see what can be learned from this.

\section{GTG and matter}\label{sec:GTG}
By virtual of GC, GTG is constructed such that gravitational effects are described by a pair of gauge fields, $\bar{h}(a)=\bar{h}(a,x)$ and $\omega(a)=\omega(a,x)$, defined over a flat Minkowski background spacetime~\citep{1998RSPTA.356..487L}, where $x$ is the STA position vector and is usually suppressed for short.

The first of them, $\bar{h}(a)$, is a position-dependent linear function mapping the vector argument $a$ to vectors. The introduction of $\bar{h}(a)$ ensures covariance of the equations under arbitrary local displacements (or an arbitrary remapping $x'=f(x)$) of the matter fields in the background spacetime. In order to understand the physical meaning of the $\bar{h}$ field, we first define the covariant displacement transformation as
\begin{equation}\label{eq:covariant displacement}
M(x) \xrightarrow{x'=f(x)} M'(x)=M(x'),
\end{equation}
so that the equations $A(x)=B(x)$ and $A(x')=B(x')$ have exactly the same physical content. Suppose we have a vector field $b(x)=\nabla\phi(x)$, where $\phi(x)$ is a scalar field that is already covariant under displacement, i.e., $\phi'(x)=\phi(x')$. Now can we write $b'(x)=b(x')$ or $\nabla\phi'(x)=\nabla_{x'}\phi(x')$? By the chain rule we find
\begin{equation}\label{eq:chain rul}
a\cdot\nabla\phi'(x)=a\cdot\nabla\phi(x')=(a\cdot\nabla f(x))\cdot\nabla_{x'}\phi(x')=\textsf{f}(a)\cdot\nabla_{x'}\phi(x')=a\cdot\bar{\textsf{f}}(\nabla_{x'})\phi(x'),
\end{equation}
where $\textsf{f}(a)=a\cdot\nabla f(x)$ is a linear function of $a$ and an arbitrary function of $x$, and $\bar{\textsf{f}}(\nabla_{x'})=\nabla f(x)\cdot\nabla_{x'}$, and we call $\bar{\textsf{f}}$ the adjoint of $\textsf{f}$, satisfying $a\cdot\textsf{f}(b)=\bar{\textsf{f}}(a)\cdot b$, or $\bar{\textsf{f}}(a)=\partial_b\langle\textsf{f}(b)a\rangle$.
It follows that $\nabla\phi'(x)=\bar{\textsf{f}}(\nabla_{x'}\phi(x'))$, or
\begin{equation}\label{eq:nabla displacement}
\nabla_x=\bar{\textsf{f}}(\nabla_{x'}) \ \ \ \ \text{and} \ \ \ \ \bar{\textsf{f}}^{-1}(\nabla_x)=\nabla_{x'},
\end{equation}
which shows us that $b(x)$ is not covariant under displacement. In order to make the objects such as $b(x)$ covariant, we must introduce a position-gauge field $\bar{\textsf{h}}(a,x)$, which is a linear function of $a$ and arbitrary function of $x$, so that
\begin{equation}\label{eq:h field}
\bar{\textsf{h}}(a,x)\xrightarrow{x'=f(x)}\bar{\textsf{h}}'(a,x)=\bar{\textsf{h}}(\bar{\textsf{f}}^{-1}(a),x').
\end{equation}
Now if we redefine $b(x)=\bar{\textsf{h}}(\nabla\phi(x))$, then
\begin{equation}\label{eq:covariant b}
b(x)=\bar{\textsf{h}}(\nabla\phi(x))\xrightarrow{x'=f(x)}b'(x)=\bar{\textsf{h}}'(\nabla\phi'(x))
=\bar{\textsf{h}}(\bar{\textsf{f}}^{-1}(\nabla\phi'(x)))=\bar{\textsf{h}}(\nabla_{x'}\phi(x'))=b(x'),
\end{equation}
which becomes covariant. The $\bar{h}(a)$ field plays the same role of vierbein in the tensor calculus approach of gauge theory of gravity~\citep{1976RevModPhys.48.393,1998RSPTA.356..487L}. We now give the relationship between $\bar{h}(a)$ and the metric tensor $g_{\mu\nu}$ in GR. We define a position gauge invariant directional derivative as~\citep{2005FoPh...35..903H}
\begin{equation}\label{eq:directional derivative}
a\cdot\bar{h}(\nabla)=\underline{h}(a)\cdot\nabla,
\end{equation}
where $\underline{h}$ is the adjoint of $\bar{h}$ defined by $\underline{h}(b)\equiv\partial_c(\bar{h}(c)\cdot b)$, and $a$ is an invariant vector (i.e., for $x'=f(x)$, $a(x)$ is transformed to $a'(x)=a(x')$). So $\underline{h}$ maps tangent vectors to tangent vectors and $\bar{h}$ maps cotangent vectors to cotangent vectors.  For a given set of coordinates \{$x^{\mu}, \mu=0, 1, 2, 3$\}, we introduce the basis vectors
\begin{equation}\label{eq:basis}
e_{\mu}\equiv\frac{\partial x}{\partial x^{\mu}}, \ \ \ e^{\mu}\equiv\nabla x^{\mu},
\end{equation}
which satisfy $e_{\mu}\cdot e^{\nu}=\delta^{\nu}_{\mu}$. From these vectors we further define vectors
\begin{equation}\label{eq:square root of metric}
g_{\mu}\equiv\underline{h}^{-1}(e_{\mu}), \ \ \ g^{\mu}\equiv\bar{h}(e^{\mu}).
\end{equation}
These vectors satisfy the relation
\begin{equation}\label{eq:orthonormal basis}
g_{\mu}\cdot g^{\nu}=\underline{h}^{-1}(e_{\mu})\cdot\bar{h}(e^{\nu})=e_{\mu}\cdot\bar{h}^{-1}(\bar{h}(e^{\nu}))=e_{\mu}\cdot e^{\nu}=\delta^{\nu}_{\mu}.
\end{equation}
The metric tensor then is given by
\begin{equation}\label{eq:metric}
  g_{\mu\nu}\equiv g_{\mu}\cdot g_{\nu}.
\end{equation}

Let $x(\tau)$ be a time like curve (where $\tau$ is the proper time), a mapping $f: x\rightarrow x'=f(x)$ induce the transformation
\begin{equation}\label{eq:map velocity}
\dot{x}=\frac{d x}{d\tau}\rightarrow \dot{x}'=\frac{d x'}{d\tau}=\frac{d x}{d\tau}\cdot\nabla f(x).
\end{equation}
Comparing (\ref{eq:map velocity}) with (\ref{eq:directional derivative}), we introduce an invariant velocity $v=v(x(\tau))$ as~\citep{2005FoPh...35..903H}
\begin{equation}\label{eq:invariant velocity}
\dot{x}=\underline{h}(v), \ \ \ v=\underline{h}^{-1}(\dot{x}).
\end{equation}
From the known formula $dx=dx^{\mu}e_{\mu}$, the invariant normalization $v^2=1$ induces the invariant line element on a timelike curve in GR
\begin{equation}\label{eq:line element}
d\tau^2=[\underline{h}^{-1}(dx)]^2=g_{\mu\nu}dx^{\mu}dx^{\nu}.
\end{equation}

Another gauge field, $\omega(a)$, is a position-dependent linear function mapping the vector argument $a$ to bivectors. Its introduction ensures covariance of the equations of physics under local Lorentz rotations.  Under local Lorentz rotations, the multivector $M$ transforms as $M'=RM\tilde{R}$ and the spinor $\psi$ transforms as $\psi'=R\psi$. To ensure covariance of the quantities like $\bar{h}(\nabla)M$ and $\bar{h}(\nabla)\psi$ under local Lorentz rotations, $\bar{h}(\nabla)$ should be replaced by the covariant derivative $D$ which is given by~\citep{1998RSPTA.356..487L}
\begin{equation}\label{eq:covariant derivative on M}
DM\equiv\partial_a a\cdot DM=\partial_a\left(a\cdot\bar{h}(\nabla)M+\omega(a)\times M\right),
\end{equation}
and
\begin{equation}\label{eq:covariant derivative on psi}
D\psi\equiv\partial_a\left(a\cdot\bar{h}(\nabla)\psi+\frac{1}{2}\omega(a)\psi\right).
\end{equation}

The field strength corresponding to the $\omega(a)$ gauge field is defined by
\begin{equation}\label{eq:curvature}
R(a\wedge b)\equiv L_a\omega(b)-L_b\omega(a)+\omega(a)\times\omega(b),
\end{equation}
where $L_a\equiv a\cdot\bar{h}(\nabla)$ and $a$, $b$ are constant vectors. Ricci tensor $R(a)$, Ricci scalar $R$ and Einstein tensor $G(a)$ are defined, respectively, as:
\begin{eqnarray}
  R(a) &=& \partial_b\cdot R(b\wedge a), \\
  R &=& \partial_a\cdot R(a), \\
  G(a) &=& R(a)-\frac{1}{2}aR.
\end{eqnarray}

The overall action integral is of the form
\begin{equation}\label{eq:action}
S=\int|d^4x|\det(\textsf{h})^{-1}(\frac{1}{2}R-\kappa\mathcal{L}_m),
\end{equation}
where $\mathcal{L}_m$ describes the matter content and $\kappa=8\pi$. As mentioned at the end of Appendix \ref{sec:STA and Dirac}, due to lacking of appropriate differential equations governing the spinor field for macro bodies, we temporally ``borrow" $\mathcal{L}_m$ from Dirac theory of electrons, which has been well-studied \citep{1998RSPTA.356..487L,1998JMP....39.3303D,1997GReGr..29.1527C}. From (\ref{eq:action}) we obtain the following equations describe the field coupled self-consistently to gravity~\citep{1998JMP....39.3303D}:
\begin{eqnarray}
  \text{torsion:}\quad   D\wedge\bar{h}(a) &=& \kappa \bar{h}(a)\cdot S, \label{eq:torsion} \\
  \text{Einstein:}\qquad  G(a) &=& \kappa T(a) \label{eq:Einstein equation} \\
  \text{Dirac:}\qquad    D\psi i\sigma_3 &=& m\psi\gamma_0, \label{eq:Dirac equation}
\end{eqnarray}
where $D\wedge\bar{h}(a)$ is the gravitational torsion which is determined by the matter spin density, $S\equiv\frac{1}{2}\psi i\gamma_3\tilde{\psi}$ is the spin trivector, $\kappa=8\pi$, and
\begin{equation}\label{eq:energy-momentum tensor}
T(a)=\langle a\cdot D\psi i\gamma_3\tilde{\psi}\rangle_1
\end{equation}
is the matter stress-energy tensor. We can solve (\ref{eq:torsion}) for $\omega(a)$ to obtain~\citep{1998JMP....39.3303D}
\begin{equation}\label{eq:omega}
  \omega(a)=\omega'(a)+\frac{1}{2}\kappa a\cdot S=-H(a)+\frac{1}{2}a\cdot[\partial_b\wedge H(b)]+\frac{1}{2}\kappa a\cdot S,
\end{equation}
this defines $\omega'(a)$ as the $\omega$-function in the absence of torsion, and
\begin{equation}\label{eq:H function}
H(a)\equiv \bar{h}(\dot{\nabla}\wedge\dot{\bar{h}}^{-1}(a))=-\bar{h}(\dot{\nabla})\wedge\dot{\bar{h}}(\bar{h}^{-1}(a)),
\end{equation}
where `overdot' notation is employed to denote the scope of a differential operator. It proves convenient if we employ the primed symbols to denote the torsion free part of the curvature tensors, we obtain~\citep{1998JMP....39.3303D}
\begin{eqnarray}
  R(a\wedge b) &=& R'(a\wedge b)+\frac{1}{4}\kappa^2[(a\wedge b)\cdot S]\cdot S \nonumber \\
               & &-\frac{1}{2}\kappa[(a\wedge b)\cdot D]\cdot S, \\
  R(a) &=& R'(a)+\frac{1}{2}\kappa^2(a\cdot S)\cdot S \nonumber \\
       & &-\frac{1}{2}\kappa a\cdot(D\cdot S), \\
  R &=& R'+\frac{3}{2}\kappa^2S^2.
\end{eqnarray}

%\begin{thebibliography}{}
%\bibitem{Hestenes:1966}
%  D.~Hestenes, spacetime algebra (Gordon and Breach, New York, 1966).

%\bibitem{Hestenes:1986}
%D.~Hestenes, and G.~Sobczyk, Clifford Algebra to Geometric Calculus, a Unified Language for Mathematics and Physics (Kluwer Academic, Dordrecht, 1986).

%\bibitem{Hestenes:1999}
%D.~Hestenes, New Foundations for Classical Mechanics (Kluwer Academic, Dordrecht, 2nd ed., 1999).

%\bibitem{Hestenes:2003a}
%  D.~Hestenes, Am.\ J.\ Phys. {\bf 71}, 104 (2003).

%\bibitem{Hestenes:2003b}
% D.~Hestenes, Am.\ J.\ Phys. {\bf 71}, 691 (2003).

% \bibitem{Doran:2003}
% C.~Doran, and A.~Lasenby, Geometric Algebra for Physicists (Cambridge University Press, Cambridge, 2003).

% \bibitem{Lasenby:1998}
% A.~Lasenby, C.~Doran, and S.~Gull, Phil. Trans. R. Lond. A {\bf 356}, 487 (1998).

% \bibitem{Hehl:1976}
% F.~W.~Hehl, P.~von der Heyde, G.~D.~Kerlick, and J.~M.~Nester, Rev.~Mod.~Phys. {\bf 48}, 393 (1976).

% \bibitem{Hestenes:2005}
% D.~Hestenes, Foundations of Physics {\bf 35}, 903 (2005).

% \bibitem{Doran:1998}
% C.~Doran, A.~Lasenby, A.~Challinor, and S.~Gull, J.~Math.~Phys. 39(6), 3303 (1998).

%\end{thebibliography}

\end{document}